# Multicolor Graphdiyne Random Lasers

*Xiantao Jiang[1, 3, †], Xuemei Zhao[1, †], Wenli Bao[1, †], Rongchao Shi[2], Jinlai Zhao[1], Jianlong Kang[1], Xuefeng Xia[1], Hualong Chen[1], Hongbo Li[1], Jialiang Xu[2], and Han Zhang[1*]*

[1]Shenzhen Engineering Laboratory of Phosphorene and Optoelectronics, Collaborative Innovation Center for Optoelectronic Science and Technology, College of Optoelectronic Engineering, Shenzhen University, Shenzhen 518060, China

[2]School of Materials Science and Engineering, National Institute for Advanced Materials, Nankai University, Tongyan Road 38, Tianjin 300350, China

[3]College of Chemistry and Environmental Engineering, Shenzhen University, Shenzhen, 518060, China

[†]These authors contributed equally.

*Corresponding to: hzhang@szu.edu.cn

**Abstract:** By breaking the restriction of mirrors, random lasers from a disordered medium have found unique applications spanning from displays, spectroscopy, biomedical treatments, to Li-Fi. Gain media in the form of two-dimension with distinct physical and chemical properties may lead to the next-generation of random lasers. Graphdiyne, a 2D graphene allotrope with intrigued carbon hybridization, atomic lattice, and optoelectronic properties, has attracted increasing attention recently. Herein, the photon emission characteristics and photo-carrier dynamics in graphdiyne are systematically studied, and the multicolor random lasers have been unprecedently realized using graphdiyne nanosheets as the gain. Considering the well bio-compatibility of graphdiyne, these results may look ahead a plethora of potential applications in the nanotechnology platform based on graphdiyne.

**Introduction**

Coherent laser source, delivered by a disordered gain medium [1] in the absence of mirrors, has found big need gaps remained though the traditional solid-state lasers have been thriving for years. Compared to regular lasers, the angular distribution of a random laser can fully cover the solid angle of 4π, which perfectly meets the requirement of display applications. Moreover, the disordered gain medium can be fabricated low-costly and be painted on arbitrarily shaped substrates [2]. Injection of the small gain medium to a particular biological tissue can distinguish themselves with distinct emission properties, e.g., spectrum and lifetime, which leads to the utilization for bio-imaging and diagnosis [3]. The limitations of laser modes and frequency due to the 2π-phase mirror rebounding in regular lasers have been broken in random ones. In this scenario, the light experiences multiple scattering in the systems that enable the spectrum blank filling [4] and speckle-free imaging [5].

Semiconductor nanosheets and quantum dots like CdSe [6], PbSe [7], ZnO [8] and perovskites [9] have been vividly investigated as the efficient gain medium. Yet, the biomedical toxicity, narrow tuning of optical frequency and intrinsic instability of materials remains a hurdle to overcome. Two-dimensional (2D) materials with extraordinary optoelectronic properties like atomic-layer-depended energy structure and high emission yields have recently been receiving increased attention [10-16]. Among them, graphdiyne (GDY), a novel 2D graphene allotrope with unique sp–sp$^2$ carbon hybridization, uniform pores, highly p-conjugated structure, and high bio-compatibility, has set up a new wave of investigation in carbon-based nanotechnologies in the recent decade [17-20]. Up to date, GDY has been intensively studied in various application fields, including catalysis[21], energy storage [22], sensing [23], and biomedical treatments [24]. Different from gapless graphene, single layer GDY is ascribed to be a narrow bandgap semiconductor, with bandgap around 0.44 to 1.47 depending on the calculation methods[18]. Furthermore, the energy structures can be largely tuned by the number of layers, stacking symmetric, strains, and chemical decorations [18, 25], endowed GDY with colorful optical features [26, 27]. Giving great potential as a novel random laser gain medium, the photon emission properties of GDY has just entered the field of view of researchers [25, 28, 29]. Herein, large-size graphdiyne was synthesized and subsequently used to prepared homogenous nanosheets dispersing in chlorobenzene solvent. The photoemission properties and photo-carrier dynamics in this solution were systematically investigated. Multi-color random laser from 450 nm to 700 nm based on GDY nanosheets has been unprecedentedly demonstrated, suggesting the promising use of GDY for white color displays, biomedical diagnose and imaging.

**Results**

The synthesis procedures of graphdiyne followed the procedures reported in Refs. [17, 30-32], as shown in **Figure 1a** (more details can be found in the Experimental section). Large size (a few centimeters) graphdiyne nanosheet can be obtained as a light yellow film on quartz via wet stripping in the FeCl$_3$ solution (Figure 1b). Figure 1c shows the high-resolution transmission electron microscope (HR-TEM) of GDY, in which the lattice fringes with an interval of 0.294 nm are clearly shown, in good agreement with previous reports [17]. The selected area electron diffraction (SAED) pattern as shown in Figure 1c (inset) confirms the nature of the high crystallinity of the prepared GDY. Raman spectra can be used to evaluate the uniformity of GDY (Figure 1g). Four Raman peaks were observed, where the peaks at 1378 and 1575 $cm^{-1}$ are ascribed to the D and G band, and 1920 and 2168 $cm^{-1}$ originates from the vibration of acetylenic linkages [17]. X-ray photoelectron spectroscopy (XPS) is employed to analyze the components of GDY. The four subpeaks rooted from C1s at 284.4, 284.9, 286.2, and 288.5 eV represents the orbitals in C-C ($sp^2$), C-C (sp), C-O, and C=O bonds, respectively, and the area ratio of $sp/sp^2$ is 2 which confirms the effective linkage of benzene ring and two acetylenic groups (Figure 1g). The solution of GDY nanosheets can be obtained after ultrasonication treatment for 12 hours. The particle thickness and size are presented in Figure 1e and 1f. The thickness of GDY nanosheets is well below 65 nm, and the determined average particle size is 389 nm.

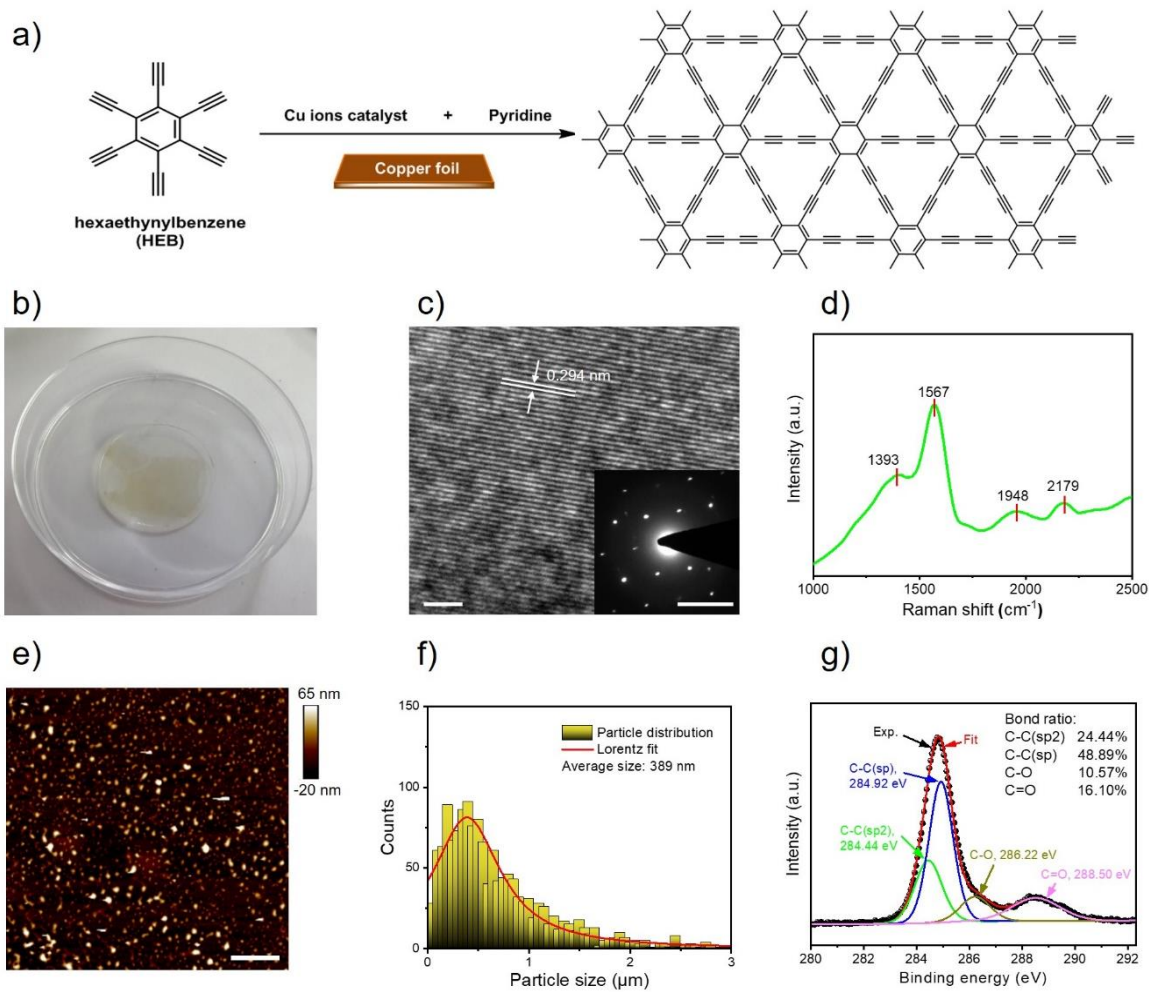

**Figure 1 Morphology characterizations of graphdiyne. a)** The synthetic scheme of the graphdiyne. **b)** Photograph of the prepared large size GDY sheets on a quartz via wet stripping. **c)** HR-TEM image of graphdiyne, scale bar: 2 nm. Inset shows the corresponding SAED pattern, scale bar: 5 1/nm. **d)** Raman spectra of the as-prepared GDY film. **e)** AFM image of the GDY nanosheet showing the particle thickness. Scale bar: 10 μm. **f)** The particle size distribution in **e)**. The average particle size is 389 nm. **g)** XPS spectra of graphdiyne nanosheet for C1s.

The linear optical absorption spectrum of GDY nanosheets is presented in Figure 2a, in which three weak absorption shoulders peak at 260, 318 and 367 nm can be observed from the broad absorption background. The prepared GDY solution is highly transparent for wavelength longer than 420 nm, which is in agreement with the photoluminescence excitation (PLE) spectra as shown in Figure 2c. The photoluminescence was characterized under the excitation of different wavelengths ranging from 280 to 400 nm (see Figure 2b and Figure S2) that well cover the absorption bands. The obtained emission bands are independent of the excitation wavelengths, which is different from previous reports in fluorinated graphdiyne [25]. The emission blue/red shift upon the excitation frequency is normally observed in other 2D materials and highly depends on the particle size and micro-environment. The observed excitation wavelength independence of emission can be ascribed to the even particle size distribution and high homologous of the solution. The maximum emission intensity was achieved under the excitation of 360 nm. The three emission peaks (407, 430 and 455 nm) are found to share a highly similar PLE spectrum and excitation intensity normalized by their corresponding emission intensity (see Figure S3). This indicates these three peaks experienced the same energy transitions in GDY. This conclusion is further confirmed by their photoluminescence lifetime characterizations as shown in Figure 2d. All the three peaks feature a lifetime of ~1.6 ns. There is no emission lifetime of GDY has been reported previously. However, this value is a bit shorter than other carbon-based 2D materials or quantum dots, like carbon nanodots (5~9 ns) [11] and graphene quantum dots (3~5 ns) [10], indicating a smaller gain and challenge of laser realization.

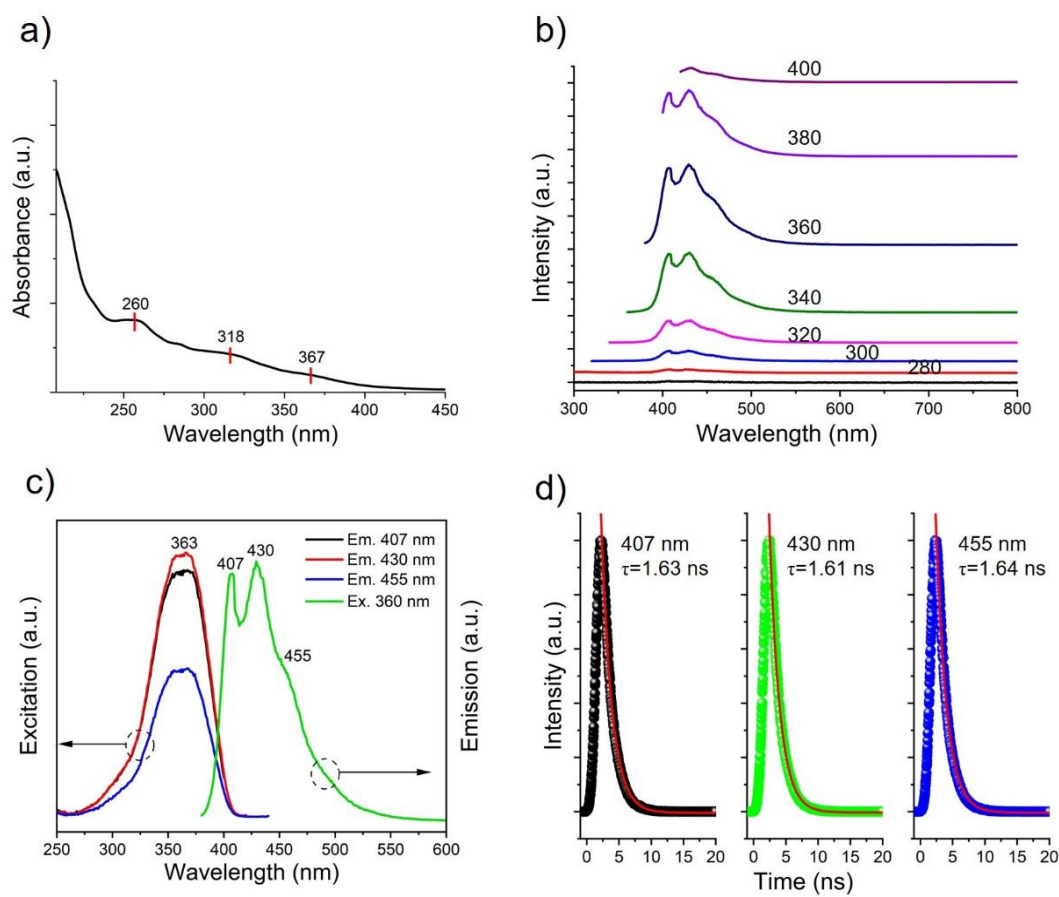

**Figure 2 Optical properties of graphdiyne. a)** Linear absorption spectrum of the GDY nanosheets. **b)** Emission spectra of GDY nanosheets under different excitation wavelengths. **c)** Emission spectrum (green) under the excitation of 360 nm, with the peak positions located at 407 nm, 430 nm, and 455 nm. Excitation spectra of emission peak at 407 nm, 430 nm, and 455 nm, the peak position is 363 nm. **d)** Photoluminescence lifetime of 407 nm, 430 nm and 455 nm under the excitation of 360 nm. The lifetimes are determined to be 1.63 ns, 1.61 ns, and 1.64 ns, respectively.

To investigate the photo-carrier dynamics in the visible regime in GDY nanosheets solution. A transient absorption (TA) spectrometer is applied with a time resolution of ~90 fs. The excitation wavelength is 400 nm, and the probe wavelength spans from 440 to 780 nm. The experimental results are shown in Figure 3a, where a strong excited-state-absorption (ESA) band can be observed from 450 nm to 600 nm. The deep blue color below 450 nm is due to the ground-state-bleaching (GSB) of the pump light. Global fitting of the experimental results delivered two principle TA spectrum and two principle kinetics as shown in Figure 3b and 3c. The principle ESA spectrum features three main peaks located at 461 nm, 510 nm and 530 nm, which show good agreement with the following laser experiments (Figure 4), though ~380 meV stokes shift can be resolved compared to the steady emission peaks (see Figure 2b). The principle kinetics cannot be suitable fit via a single decay process. A two exponential decay components process fitting give the lifetime of $\tau_1$ = 2.98 ps, $\tau_2$ = 140.4 ps for kinetics 1, and $\tau_1$ = 12.4 ps, $\tau_2$ = 180.1 ps for kinetics 2, as shown in Figure 3c. The ESA spectrum at different delay time is shown in Figure S4. The proposed two steps transitions are presented in the inset of Figure 3a. The ground-state carries can be elevated to excited-states under the pump of 400 nm (~3.1 eV) photons, and then quickly thermalized to an intermediate state on the time scale of 3~12 ps. Considering the multi-peaks of ESA and emission (Figure 4), sub-level fine structures can exist in the intermediate state (inset of Figure 3a). Later on, the excited carriers will return to ground-states at the price of photon emission around ~2.38 eV on the time scale of 140~180 ps. Note that the lifetime of the TA spectrum is different from the steady PL processes where no relaxation probe light is presented.

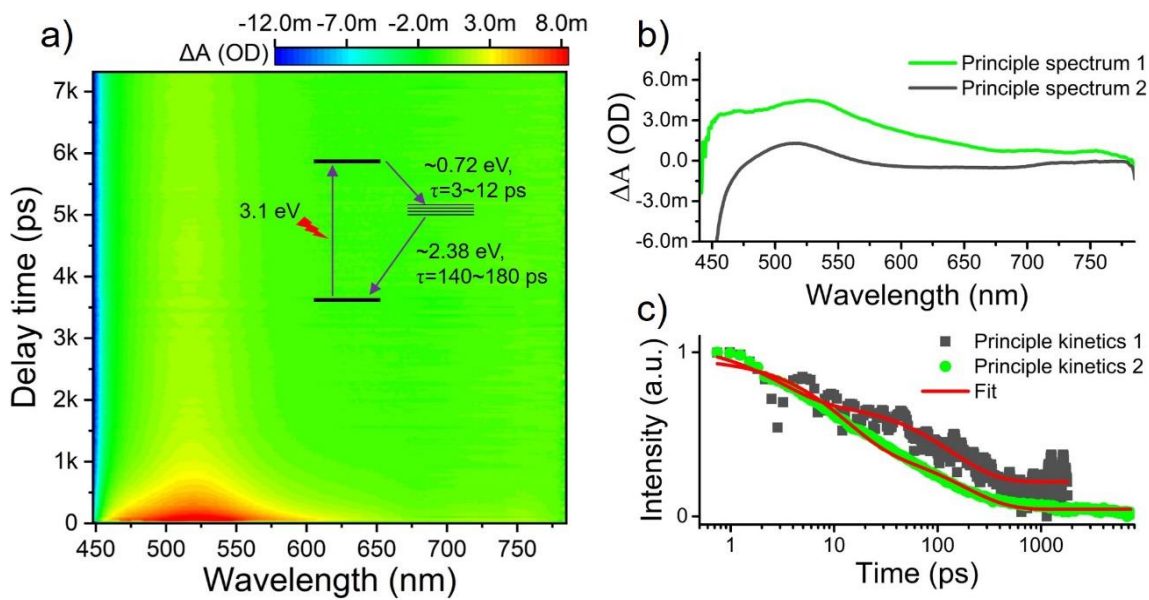

**Figure 3 Photo-carrier dynamics in graphdiyne nanosheets. a)** Map of experimental transient absorption characterization of GDY nanosheets solution. The inset shows the proposed photo-carrier transition processes. **b)** Global fitted spectra and **c)** kinetics of the experimental results.

The random laser performance of the GDY solution is presented in Figure 4. Three available excitation wavelengths (390, 400 and 410 nm) which are close to the effective PLE peak (see Figure 2c) are used as the pump source. As the pump intensity increases, sharp emission spikes emerged on a broad background, which indicates the successful generation of random lasers. It could see that the random laser threshold increases as a function of the pump wavelength. Note, even at low pump intensities, photoluminescence also can be well observed at all the pump wavelengths, see Figure S6-S23. The integrated intensity at the three pump wavelengths as a function of the pump intensity is shown in Figure 5a-c. The random laser thresholds for the three pump wavelengths are determined to be 7.3, 16.4 and 23.0 kW/cm$^2$. This sequence is consistent with the previous PLE spectrum, where the efficient excitation decreases as these wavelength increases (Figure 2c). Since the effective excitation above 400 nm is insignificant, it requires much higher excitation intensity to achieve the laser threshold. The spectra of this GDY-based random laser covers most of the visible region from 450 nm to 700 nm (see more in Figure S6-S23). The chromaticity of the random laser spectrum under different pump wavelengths and intensities is summarized in Figure 5d, corresponding to the spectra in Figure 4. It is shown the color of the random laser can be largely turned from light green to light red through engineering the pump parameters. An image of the GDY random laser under testing is shown in Figure 5e.

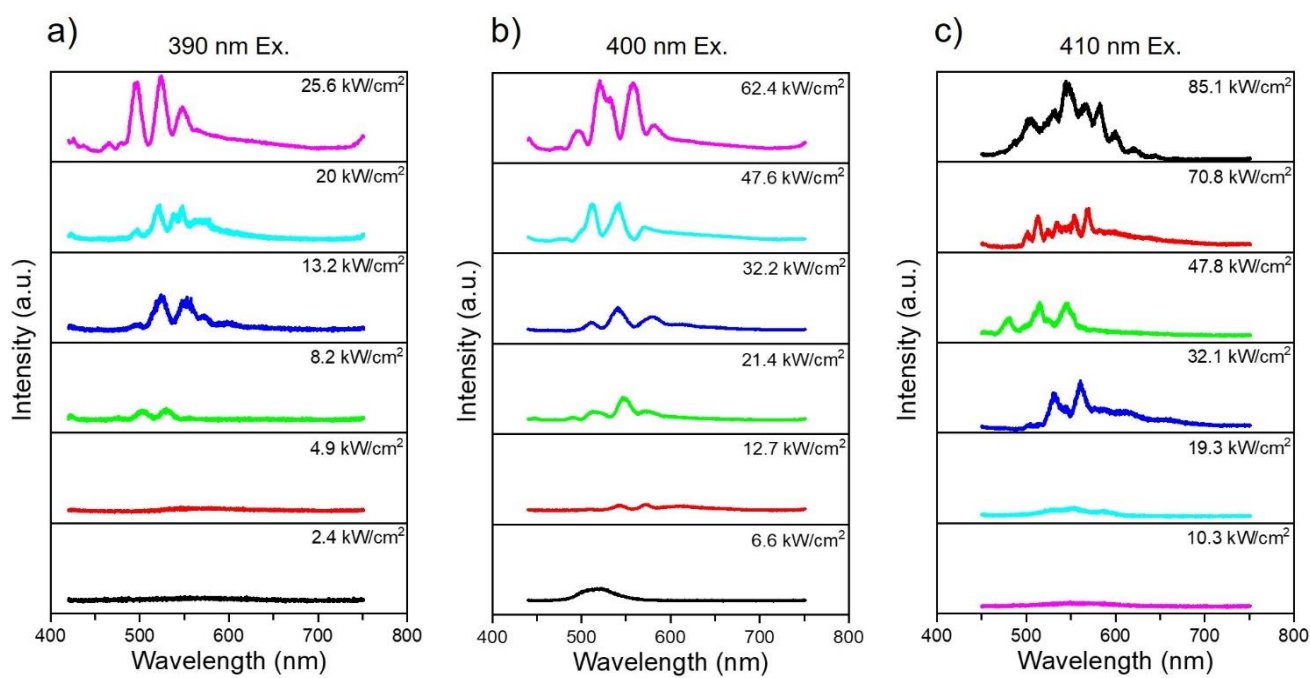

**Figure 4. Random laser performance of graphdiyne nanosheets.** The typical emission spectra of graphdiyne nanosheets solution under the excitation of **a)** 390, **b)** 400, and **c)** 410 nm. The pump intensity is shown in the insets of the figures. The random laser emerged when the pump power up to 8.2, 12.7 and 32.1 kW/cm$^2$ for the three excitation wavelengths, respectively.

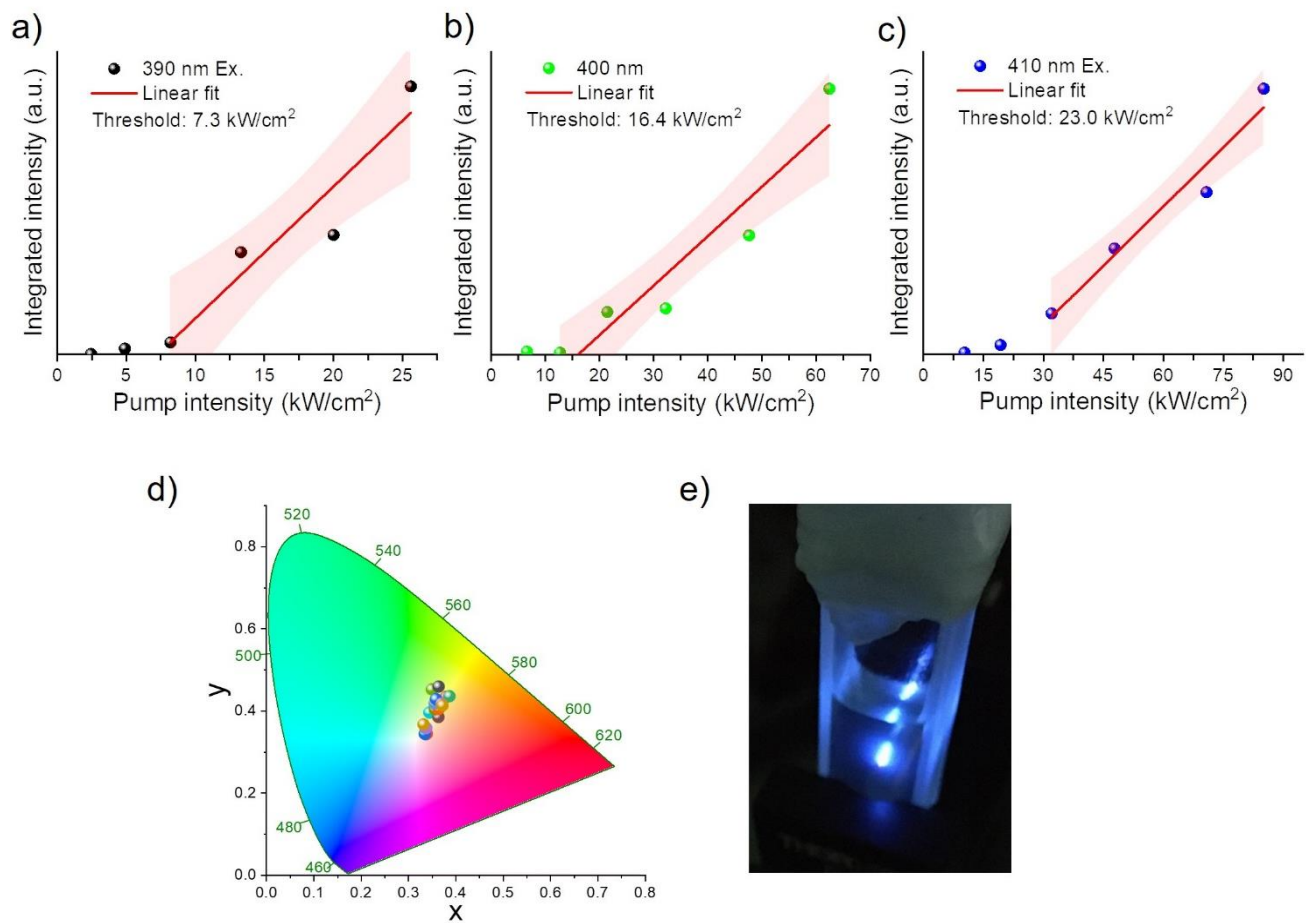

**Figure 5** The integrated emission intensity as a function of pump wavelength at **a)** 390 nm, **b)** 400 nm, and **c)** 410 nm, respectively. The fitted random laser thresholds are shown in the figures. **d)** The 1931 CIE color space chromaticity diagram of the multi-color GDY random lasers, corresponding to the emission spectra in Figure 4. **e)** Photoimage of the GDY random laser under testing.

According to the theory of the random laser[33], the active gain medium (here is the GDY nanosheets) will randomize and amplify the emitted photons in the solution. The pre-condition for a random laser is the photon amplification is larger than the optical losses in the random nanostructures. Since the particle size of GDY is around 389 nm, which is smaller than the wavelength, Mie scattering cannot offer effective optical feedback for the generation of the random laser. Localized bubbles under the intense excitation are proposed to realize effective feedback due to the strong photothermal properties of GDY [24, 34]. The low surface tension of the solution is in favor of large bubble formation according to the derived Rayleigh-Plesset equation [15, 35], and thus high strength of light scattering. The surface tension of the chlorobenzene is 33.6 mN/m@20 °C, which is smaller than that of water (72.8mN/m @20 °C), suggesting the easier bubble formation. In this experiment, both the temporal and spectral chaotic behavior [1] of random laser have been observed. The emission peaks and intensity of random laser varies at different pump shots (Figure S6-S23). Brown motions of the gain medium, instabilities of the generated bubbles changes as the time goes, and the high freedom of modes competition due to the multi-scattering in the system give the spectral diversity.

**Conclusions**

To conclude, large size graphdiyne film was prepared via a feasible wet stripping method. Reducing the particle size to a few tens nanometers, GDY nanosheets have been shown excellent photon emission performance in the visible region. The photon transitions obey a three-level energy structure scheme and show a broadband emission covering 450-700 nm. Photons with a wavelength shorter than 400 nm are found to be an efficient excitation source. Using different excitation wavelengths and intensities, multicolor random lasers from GDY nanosheets solution have been unambiguously demonstrated. The chaotic characteristics of random laser originate from Brown motions, instabilities of bubbles, mode competitions ignored in most of the previous reports are discussed. Up to date, the optical properties of graphdyine is far from beening well explored, this letter may shed light on the foreseeable coming of graphdiyne enabled nanophotonics.

**Experimental Section**

*Graphdiyne synthesis*

The procedures of the synthesis experiments are presented in Figure 1a. Deprotection of the hexakis[(trimethylsilyl)ethynyl]benzene with the tetrabutylammonium fluoride (TBAF) solution for 10 min at 8°C to afford the monomer of hexaethynylbenzene in good yield (62%). The graphdiyne was smoothly grown on the surface of copper foil in the presence of pyridine by the Suzuki cross-coupling reaction of the monomer of hexaethynylbenzene for 72 h at 60°C under nitrogen atmosphere. The copper foil is not only the catalyst for the cross-coupling reaction but also is the substrate for growing graphdiyne film [30]. GDY nanosheets were prepared by evenly mixing GDY power in chlorobenzene (CB) at a concentration of 0.5 mg ml$^{-1}$ and then ultrasonic treatment at 60°C for 12 hours. The resulted mixture was then centrifuged at 8000 rpm for three minutes to remove the precipitate. The GDY nanosheets were suspended in the supernatant with a concentration of ~0.05 mg ml$^{-1}$.

*Morphology characterizations*

Atomic Force Microscope measurements were conducted using the Dimension ICON from Bruker and the sample was prepared by dropping the solution on the substrate of mica. UV-vis absorption spectra were recorded on the Agilent Cary 60 spectrophotometer. Transmission electron microscopy (TEM) and selected area electron diffraction (SAED) pattern measurements were conducted with FEI Tecnai G2 f20 s-twin 200kV transmission electron microscopes using an accelerating rate voltage of 120 keV. The X-ray photoelectron spectrometer (XPS) was collected on the VG Scientific ESCALab250Xi X-Ray photoelectron spectrometer, using Al Kα radiation as the excitation sources. The banding energies obtained in the XPS analysis were corrected regarding C1s (284.8 eV). Raman spectra were taken on a Horiba LabRAM HR800 Raman spectrometer at a resolution of 1.5 cm$^{-1}$ by using the 532 nm line of an Argon ion laser as the excitation source.

*Transient absorption microscope*

The carrier dynamics in GDY solution in the visible region were characterized via a commercial transient absorption spectrometer (Helios fire, Ultrafast Systems, USA). The pump source was delivered by an OPA system at 400 nm. A broadband probe wavelength band was generated through a CaF$_2$ crystal under the excitation of a femtosecond Ti: sapphire oscillator. The resolution of the delay line is 15 femtosecond, and the resolution of the pulsed laser sources is 90 femtosecond. The spacial resolution of the pump/probe beam is around 50 micrometers. The GDY solution was dispersed in the IPA solution and then hold in a 10 mm thick cuvette for carrier dynamics measurement.

*Random laser*

The pump light (390, 400 and 410 nm) is delivered by an OPA system with a repetition of 1 kHz and a pulse duration of ~150 fs. The pump light is focused by a plano-convex lens with a focus lens of 30 mm. The GDY nanosheets solution is held by a 5 mm thick cuvette and placed on the focus. The intensity of the pump is manually altered by an ND filter (NDC-50S-3M, Thorlabs). The pump intensities at different pump wavelengths are shown in Figure S5. The emission is collected by a fiber-coupled high-resolution spectrometer (iHR550, Horiba).

**Author Contributions**

The manuscript was written through contributions of all authors. All authors have given approval to the final version of the manuscript. †These authors contributed equally.

**Funding Sources**

This work is supported by the State Key Research Development Program of China (Grant No. 2019YFB2203503), National Natural Science Fund (Grant Nos. 61805146, 61875138, 61961136001, 21773168, and U1801254), and Science and Technology Innovation Commission of Shenzhen (KQTD2015032416270385, JCYJ20170811093453105, JCYJ20180307164612205, and GJHZ20180928160209731). Authors also acknowledge the support from Instrumental Analysis Center of Shenzhen University (Xili Campus).

**References**

1. Wiersma, D. S., The physics and applications of random lasers. *Nat. Phys.* **2008,** *4* (5), 359-367.

2. Cao, H. J. W. i. r. m., Lasing in random media. **2003,** *13* (3), R1-R39.

3. Carvalho, M.; Lotay, A.; Kenny, F.; Girkin, J.; Gomes, A. S., *Random laser illumination: an ideal source for biomedical polarization imaging?* SPIE: 2016; Vol. 9701.

4. Chang, S.-W.; Liao, W.-C.; Liao, Y.-M.; Lin, H.-I.; Lin, H.-Y.; Lin, W.-J.; Lin, S.-Y.; Perumal, P.; Haider, G.; Tai, C.-T.; Shen, K.-C.; Chang, C.-H.; Huang, Y.-F.; Lin, T.-Y.; Chen, Y.-F., A White Random Laser. *Sci Rep* **2018,** *8* (1), 2720.

5. Redding, B.; Choma, M. A.; Cao, H., Speckle-free laser imaging using random laser illumination. *Nat. Photonics* **2012,** *6* (6), 355-359.

6.  Chen, Y.; Herrnsdorf, J.; Guilhabert, B.; Zhang, Y.; Watson, I. M.; Gu, E.; Laurand, N.; Dawson, M. D., Colloidal quantum dot random laser. *Opt. Express* **2011,** *19* (4), 2996-3003.

7.  Yang, J.; Heo, J.; Zhu, T.; Xu, J.; Topolancik, J.; Vollmer, F.; Ilic, R.; Bhattacharya, P., Enhanced photoluminescence from embedded PbSe colloidal quantum dots in silicon-based random photonic crystal microcavities. **2008,** *92* (26), 261110.

8.  Lee, Y.-J.; Yeh, T.-W.; Yang, Z.-P.; Yao, Y.-C.; Chang, C.-Y.; Tsai, M.-T.; Sheu, J.-K., A curvature-tunable random laser. *Nanoscale* **2019,** *11* (8), 3534-3545.

9.  Safdar, A.; Wang, Y.; Krauss, T. F., Random lasing in uniform perovskite thin films. *Opt. Express* **2018,** *26* (2), A75-A84.

10. Zhao, S.; Lavie, J.; Rondin, L.; Orcin-Chaix, L.; Diederichs, C.; Roussignol, P.; Chassagneux, Y.; Voisin, C.; Müllen, K.; Narita, A.; Campidelli, S.; Lauret, J.-S., Single photon emission from graphene quantum dots at room temperature. *Nature Communications* **2018,** *9* (1), 3470.

11. Zhang, W. F.; Zhu, H.; Yu, S. F.; Yang, H. Y., Observation of Lasing Emission from Carbon Nanodots in Organic Solvents. **2012,** *24* (17), 2263-2267.

12. Yuan, F.; Xi, Z.; Shi, X.; Li, Y.; Li, X.; Wang, Z.; Fan, L.; Yang, S., Ultrastable and Low-Threshold Random Lasing from Narrow-Bandwidth-Emission Triangular Carbon Quantum Dots. **2019,** *7* (2), 1801202.

13. Roy, P. K.; Haider, G.; Lin, H.-I.; Liao, Y.-M.; Lu, C.-H.; Chen, K.-H.; Chen, L.-C.; Shih, W.-H.; Liang, C.-T.; Chen, Y.-F., Multicolor Ultralow-Threshold Random Laser Assisted by Vertical-Graphene Network. **2018,** *6* (16), 1800382.

14. O'Brien, S. A.; Harvey, A.; Griffin, A.; Donnelly, T.; Mulcahy, D.; Coleman, J. N.; Donegan, J. F.; McCloskey, D., Light scattering and random lasing in aqueous suspensions of hexagonal boron nitride nanoflakes. *Nanotechnology* **2017,** *28* (47), 47LT02.

15. Huang, D.; Xie, Y.; Lu, D.; Wang, Z.; Wang, J.; Yu, H.; Zhang, H., Demonstration of a White Laser with V2C MXene-Based Quantum Dots. *0* (0), 1901117.

16. Hu, H.-W.; Haider, G.; Liao, Y.-M.; Roy, P. K.; Ravindranath, R.; Chang, H.-T.; Lu, C.-H.; Tseng, C.-Y.; Lin, T.-Y.; Shih, W.-H.; Chen, Y.-F., Wrinkled 2D Materials: A Versatile Platform for Low-Threshold Stretchable Random Lasers. **2017,** *29* (43), 1703549.

17. Li, G.; Li, Y.; Liu, H.; Guo, Y.; Li, Y.; Zhu, D., Architecture of graphdiyne nanoscale films. *Chem. Commun.* **2010,** *46* (19), 3256-3258.

18. Gao, X.; Liu, H.; Wang, D.; Zhang, J., Graphdiyne: synthesis, properties, and applications. *Chemical Society Reviews* **2019,** *48* (3), 908-936.

19. Li, Y.; Liu, T.; Liu, H.; Tian, M.-Z.; Li, Y., Self-Assembly of Intramolecular Charge-Transfer Compounds into Functional Molecular Systems. *Accounts Chem. Res.* **2014,** *47* (4), 1186-1198.

20. Huang, C.; Li, Y.; Wang, N.; Xue, Y.; Zuo, Z.; Liu, H.; Li, Y., Progress in Research into 2D Graphdiyne-Based Materials. *Chemical Reviews* **2018,** *118* (16), 7744-7803.

21. Hui, L.; Xue, Y.; Yu, H.; Liu, Y.; Fang, Y.; Xing, C.; Huang, B.; Li, Y., Highly Efficient and Selective Generation of Ammonia and Hydrogen on a Graphdiyne-Based Catalyst. *Journal of the American Chemical Society* **2019,** *141* (27), 10677-10683.

22. Hwang, H. J.; Kwon, Y.; Lee, H., Thermodynamically stable calcium-decorated graphyne as a

hydrogen storage medium. *The Journal of Physical Chemistry C* **2012**, *116* (38), 20220-20224.

23. Parvin, N.; Jin, Q.; Wei, Y.; Yu, R.; Zheng, B.; Huang, L.; Zhang, Y.; Wang, L.; Zhang, H.; Gao, M.; Zhao, H.; Hu, W.; Li, Y.; Wang, D., Few-Layer Graphdiyne Nanosheets Applied for Multiplexed Real-Time DNA Detection. **2017**, *29* (18), 1606755.

24. Li, S.; Chen, Y.; Liu, H.; Wang, Y.; Liu, L.; Lv, F.; Li, Y.; Wang, S., Graphdiyne Materials as Nanotransducer for in Vivo Photoacoustic Imaging and Photothermal Therapy of Tumor. *Chemistry of Materials* **2017**, *29* (14), 6087-6094.

25. Xiao, W.; Kang, H.; Lin, Y.; Liang, M.; Li, J.; Huang, F.; Feng, Q.; Zheng, Y.; Huang, Z., Fluorinated graphdiyne as a significantly enhanced fluorescence material. *RSC Adv.* **2019**, *9* (32), 18377-18382.

26. Wu, L.; Dong, Y.; Zhao, J.; Ma, D.; Huang, W.; Zhang, Y.; Wang, Y.; Jiang, X.; Xiang, Y.; Li, J.; Feng, Y.; Xu, J.; Zhang, H., Kerr Nonlinearity in 2D Graphdiyne for Passive Photonic Diodes. **2019**, *31* (14), 1807981.

27. Guo, J.; Shi, R.; Wang, R.; Wang, Y.; Zhang, F.; Wang, C.; Chen, H.; Ma, C.; Wang, Z.; Ge, Y.; Song, Y.; Luo, Z.; Fan, D.; Jiang, X.; Xu, J.; Zhang, H., Graphdiyne-Polymer Nanocomposite as a Broadband and Robust Saturable Absorber for Ultrafast Photonics. *n/a* (n/a), 1900367.

28. Zheng, Y.-p.; Feng, Q.; Tang, N.-j.; Du, Y.-w., Synthesis and photoluminescence of graphdiyne. *New Carbon Materials* **2018**, *33* (6), 516-521.

29. Gao, X.; Zhu, Y.; Yi, D.; Zhou, J.; Zhang, S.; Yin, C.; Ding, F.; Zhang, S.; Yi, X.; Wang, J.; Tong, L.; Han, Y.; Liu, Z.; Zhang, J., Ultrathin graphdiyne film on graphene through solution-phase van der Waals epitaxy. **2018,** *4* (7), eaat6378.

30. Hebert, N.; Beck, A.; Lennox, R. B.; Just, G., A new reagent for the removal of the 4-methoxybenzyl ether: application to the synthesis of unusual macrocyclic and bolaform phosphatidylcholines. *The Journal of Organic Chemistry* **1992,** *57* (6), 1777-1783.

31. King, A. O.; Negishi, E.; Villani Jr, F. J.; Silveira Jr, A., A general synthesis of terminal and internal arylalkynes by the palladium-catalyzed reaction of alkynylzinc reagents with aryl halides. *The Journal of Organic Chemistry* **1978,** *43* (2), 358-360.

32. Sonoda, M.; Inaba, A.; Itahashi, K.; Tobe, Y., Synthesis of differentially substituted hexaethynylbenzenes based on tandem Sonogashira and Negishi cross-coupling reactions. *Organic letters* **2001,** *3* (15), 2419-2421.

33. Luan, F.; Gu, B.; Gomes, A. S. L.; Yong, K.-T.; Wen, S.; Prasad, P. N., Lasing in nanocomposite random media. *Nano Today* **2015,** *10* (2), 168-192.

34. Jin, J.; Guo, M.; Liu, J.; Liu, J.; Zhou, H.; Li, J.; Wang, L.; Liu, H.; Li, Y.; Zhao, Y.; Chen, C., Graphdiyne Nanosheet-Based Drug Delivery Platform for Photothermal/Chemotherapy Combination Treatment of Cancer. *ACS Applied Materials & Interfaces* **2018,** *10* (10), 8436-8442.

35. Soliman, W.; Nakano, T.; Takada, N.; Sasaki, K., Modification of Rayleigh–Plesset Theory for Reproducing Dynamics of Cavitation Bubbles in Liquid-Phase Laser Ablation. *Jpn. J. Appl. Phys.* **2010,** *49* (11), 116202.